\documentclass[12pt]{article}
\usepackage{graphicx}
\def\ltap{\raisebox{-.4ex}{\rlap{$\sim$}} \raisebox{.4ex}{$<$}}

\def\sinthl {${\rm sin}^2 \theta_W^{\ell\ {\rm eff}}$}
\def\sinthwsp {${\rm sin}^2 \theta_W\ $}
\def\sinthw {${\rm sin}^2 \theta_W$}
\def\sinthlsp {${\rm sin}^2 \theta_W^{\ell\ {\rm eff}}\ $}
\def\qx {$Q_X$}
\def\qxsp {$Q_X\ $}
\def\zp {$Z^{\prime}$}
\def\zpsp {$Z^{\prime}\ $}

\def\tx {$T_X$}
\def\txsp {$T_X\ $}

\def\zzpsp {$Z-Z^{\prime}\ $}

\def\alr {$A_{LR}$}
\def\alrsp {$A_{LR}\ $}
\def\afbb {$A_{FB}^b$}
\def\afbbsp {$A_{FB}^b\ $}

\def\afbc {$A_{FB}^c$}
\def\afbcsp {$A_{FB}^c\ $}
\def\qfb {$Q_{FB}$}

\def\afbl {$A_{FB}^{\ell}$}
\def\atau {$A_{\ell}(P_{\tau})$}

\def\mh {$m_H$}
\def\mhsp {$m_H$\ }
\def\chisqsp {$\chi^2$\ }
\def\chisq {$\chi^2$}

\def\bmlsp{$B-L\ $}
\def\zsp{$Z\ $}

\def\ntsp{$\rm NuTeV\ $}

\def\nuNsp{$\nu {\rm N}\ $}
\def\journal{\topmargin 0.0in   \oddsidemargin 0in
        \headheight 0pt \headsep 0pt
        \textwidth 6.5in 
\textheight 9in 
        \marginparwidth 1.5in
        \parindent 2em
        \parskip .5ex plus .1ex         \jot = 1.5ex}
\journal

\begin{document}
\begin{titlepage}

\noindent March 3, 2008      \hfill    LBNL-1619E\\

\begin{center}

\vskip .5in

{\large \zpsp Bosons, the NuTeV Anomaly, and the Higgs Boson Mass}

\vskip .5in

Michael S. Chanowitz

\vskip .2in

{\em Theoretical Physics Group\\
     Lawrence Berkeley National Laboratory\\
     University of California\\
     Berkeley, California 94720}
\end{center}

\vskip .25in

\begin{abstract}

Fits to the precision electroweak data that include the \ntsp
measurement are considered in family universal, anomaly free $U(1)$
extensions of the Standard Model. In data sets from which the hadronic
asymmetries are excluded, some of the \zpsp models can double the
predicted value of the Higgs boson mass, from $\sim 60$ to $\sim 120$
GeV, removing the tension with the LEP II lower bound, while also
modestly improving the \chisqsp confidence level.  The effect of the
\zpsp models on both \mhsp and the \chisqsp confidence level is
increased when the \ntsp measurment is included in the fit. Both the
original \ntsp data and a revised estimate by the PDG are considered. 

\end{abstract}

\end{titlepage}

\newpage

\renewcommand{\thepage}{\arabic{page}}
\setcounter{page}{1}

{\it To Lev Okun in honor of his 80'th birthday: a kind and gentle man
  who does physics as he lives his life, with simple honesty and
  integrity. Although I have told it before, the story of my first
  encounter with Lev bears retelling. It was at the 1976 International
  Conference on High Energy Physics in Tbilisi. Andre Sakharov was not
  allowed to register for the conference but was told that his
  presence would be tolerated if he wished to attend informally. While
  there was doubtless considerable sympathy for Sakharov among many of
  the Soviet physicists attending the conference, Okun was the only
  one I saw who dared to associate openly on the streets of Tbilisi
  with Sakharov during the meeting. In another characteristic decision,
  in the period following the fall of the Soviet Union, Lev chose to
  remain in Moscow to help preserve the marvelous school of
  theoretical physics at ITEP, when he could have accepted offers of
  more comfortable positions outside of Russia.}

\vskip .25in
{\it \noindent \underline{Introduction}}

When the precision electroweak data began to emerge from LEP, SLC, and
Fermilab, Okun and collaborators Novikov, Rozanov, and Vysotsky did a
complete and independent study of the one loop radiative corrections,
culminating in the LEPTOP program.\cite{leptop} Using LEPTOP they then
made the interesting discovery, contrary to the conventional wisdom of
the time, that a fourth generation of quarks and leptons is not
excluded by the precision data.\cite{lev4} They also found that a
fourth generation would raise the Higgs mass prediction significantly,
and that it could resolve the conflict between the SM (Standard Model)
prediction for \mhsp and the 114 GeV LEP II direct lower limit on
\mhsp that arises if the inconsistent determinations of the weak
interaction mixing angle \sinthlsp are attributed to underestimated
systematic error in the hadronic asymmetry measurements.\cite{mcmh} In
this paper I consider a class of \zpsp models that would also raise
the \mhsp predicton into the LEP II allowed region. In particular I
extend an earlier study\cite{mczpqx} of the EWWG\cite{ewwg}
(Electroweak Working Group) set of observables, to also include three
low energy measurements: the \ntsp $\nu N$ scattering
measurement,\cite{nutev} M\"oller scattering,\cite{moller} and atomic
parity violation.\cite{apv} Principally as a result of the \ntsp
measurement, the \zpsp models have a greater effect on the fits than
in the previous study: the central value of \mhsp is raised by a
factor 2 into the LEP II allowed region and the \chisqsp
confidence level is modestly improved in some cases.

\begin{table}
\begin{center}
\vskip 12pt
\begin{tabular}{c|c|cc|cc}
\hline
\hline
 &Experiment& {\bf A} & Pull &{\bf B} & Pull \\ 
\hline
$A_{LR}$ & 0.1513 (21)  & 0.1476  & 1.8& 0.1494  & 0.9  \\
$A_{FB}^l$ & 0.01714 (95) &0.01634  & 0.8&0.1674  &0.4  \\
$A_{e,\tau}$ & 0.1465 (32) & 0.1476 & -0.3& 0.1494 & -0.9 \\
$A_{FB}^b$ & 0.0992 (16) & 0.1035 & -2.7&  &   \\
$A_{FB}^c$ & 0.0707 (35) & 0.0739 & -0.9&  &  \\
$m_W$ & 80.398 (25) & 80.369 & 1.2 & 80.391 & 0.3 \\
$\Gamma_Z$ & 2495.2 (23) & 2495.7 &0.2& 2496.1 & -0.4  \\
$R_l$ & 20.767 (25) &20.743  & 1.0 &20.743  & 1.0  \\
$\sigma_h$ & 41.540 (37) & 41.477 &1.7  & 41.479 &1.7  \\
$R_b$ & 0.21629 (66) & 0.21586 &0.7& 0.21584 &0.7  \\
$R_c$ & 0.1721 (30) & 0.1722 &-0.04& 0.1722 &-0.04  \\
$A_b$ & 0.923 (20) & 0.935 &-0.6 & 0.935 &-0.6  \\
$A_c$ & 0.670 (27) &  0.668 & 0.07&  0.669 & 0.04 \\
$g_L^2$& 0.30005 (137)& 0.30396 &-2.9 & 0.30423& -3.1\\
$g_R^2$ &0.03076 (11)& 0.03009 & 0.6& 0.03004& 0.7 \\
$x_W(ee)$& 0.23339 (140)& 0.23145& 1.4& 0.23122& 1.55 \\
$x_W(Cs)$ &0.22939 (190)& 0.23145& -1.1& 0.23122& -1.0 \\
$m_t$ & 172.6 (1.4) &172.3  &0.2&172.3  &0.2  \\
$\Delta \alpha_5(m_Z)$ & 0.02758 (35) &0.02768 & -0.3&0.02754 & 0.1 \\
$\alpha_S(m_Z)$ &    &0.1186& &0.118  &  \\
\hline
$m_H$ & & 94 && 64 &\\
CL$(m_H > 114)$ & & 0.33 && 0.07 & \\
$m_H(95\%)$&&172&&124&\\
\hline
$\chi^2$/dof& & 28.4/15 && 19.0/13 & \\
CL($\chi^2)$ & & 0.02 &&0.12 & \\
\hline
\hline
\end{tabular}
\end{center}
\caption{SM fits with (A) and without (B) $A_{FB}^b$ and $A_{FB}^c$.}
\end{table}

{\it \noindent \underline {The SM Fit}}

The SM fit to the precision electroweak data shown in table 1 has two
$3\sigma$ anomalies which together reduce the \chisqsp confidence
level of the fit to 2\%.\footnote{Radiative corrections are from
  ZFITTER\cite{zfitter}, with two loop corrections to
  \sinthlsp\cite{xw2loop} and $m_W$.\cite{mw2loop}} One is the $3.2
\sigma$ discrepancy between the two most precise measurements of the
weak interaction mixing angle \sinthl, \alrsp and \afbb, and the
corresponding $3.2\sigma$ discrepancy between the three hadronic
asymmetry measurements (\afbb, \afbc, \qfb) and the three leptonic
ones (\alr, \afbl, \atau).\cite{ewwg} The second is the measurement of
the weak mixing angle, \sinthw,in \nuNsp scattering by the \ntsp
experiment.\cite{nutev}\footnote{Our fits use the \ntsp measurments of
  $g_L^2$ and $g_R^2$, which are simple functions of \sinthw. The
  anomaly is manifested in $g_L^2$, which is more sensitive to
  \sinthw.}  Either or both could be the result of statistical
fluctuations, genuine new physics, or underestimated systematic
uncertainty. If the cause is systematic error in the asymmetry
measurements, the hadronic asymmetries are by far the more likely
candidates since they have challenging experimental and QCD-related
theoretical issues in common, which have no counterparts in the
leptonic asymmetry measurements that are measured by three quite
independent and relatively simple techniques.\footnote{See
  \cite{mcfnal} for a more detailed discussion.}  Without prejudging
the validity of any of the possibilities, in this paper we explore the
consequences of the assumption that the hadronic asymmetry
measurements have underestimated systematic error. We then also
consider the fit without the hadronic asymmetry measurements, data set
B in table 1. 

As observed by Davidson {\it et al.}\cite{gambinoetal} the \ntsp
anomaly is also susceptible to theoretical systematic effects from
QCD.\footnote{Davidson {\it et al.} also considered an unmixed \zpsp
  coupled to $B - 3L_\mu$} In particular, a positive asymmetry in the
momentum carried by strange versus anti-strange quarks in the nucleon
sea would reduce the size of the anomaly. Recently the \ntsp
collaboration has found a $1.4\sigma$ indication for a positive
asymmetry.\cite{ntssbar} Taken at face value it would reduce the
anomaly from 3 to $2\sigma$, as noted by the PDG\cite{pdg} (Particle
Data Group), which quotes revised estimates for $g_L^2$ and
$g_R^2$. In this paper we consider both the original \ntsp results and
the PDG estimates.

Table 1, with the original \ntsp data, displays the SM
fit with (A) and without (B) the hadronic asymmetry measurements,
\afbbsp and \afbc. In addition to the three low energy measurements,
we include the usual EWWG data set, with two unimportant exceptions
that have a negligible effect on the results: the jet charge
asymmetry, the least precise of the six asymmetry measurements, which
we omit because of potential flavor-dependent systematic effects that
could effect the \zpsp fits, and the $W$ boson width, which with a 2\%
error is not a precision measurement in the sense of the other
measurements that typically have part per mil precision. The \ntsp
measurement is represented by $g_{L}^2$ and $g_{R}^2$, which in the SM
are given by $g_L^2= 1/2 - {\rm sin}^2\theta_W + 5/9\, {\rm
  sin}^4\theta_W$ and $g_R^2= 5/9\, {\rm sin}^4\theta_W$.  In fit A of
table 1 the $3.2\sigma$ conflict between \alrsp and \afbbsp is
manifested by their pulls, $+1.8$ and $-2.7$ respectively; together
with the $-2.9$ pull of the \ntsp measurement of $g_L^2$, they are
responsible for the poor 2\% confidence level of the fit. In fit B
with \afbbsp and \afbcsp excluded, the CL increases to 12\% but the
central value of the Higgs mass decreases to 60 GeV with only 7\% CL
to be in the LEP II allowed region above 114 GeV.

The best SM fit to data set A using the PDG estimate of $g_{L}^2$ and
$g_{R}^2$ has $\chi^2/N = 24.2/15$ implying a still marginal 6\%
confidence level. The Higgs mass is 85 GeV with an acceptable 31\% CL
for $m_H>114$ GeV. For data set B the confidence level of the SM fit,
shown in table 5, increases to a robust 35\% but the Higgs mass
prediction decreases to 58 GeV, with a 6\% confidence level for
$m_H>114$ GeV. This fit then also provides motivation to consider new
physics that can increase the \mhsp prediction.

{\it \noindent \underline{The \zpsp Model}}

A heavy \zpsp boson offers a natural solution to an unacceptably light
prediction for \mhsp because \zzpsp mixing with a heavy \zpsp shifts
the \zsp mass downward, corresponding to a positive contribution to
the $\rho$ parameter that offsets the negative contribution from an
increase in the Higgs boson mass.  There is a vast literature on \zpsp
models.\cite{pl} Here we consider the highly constrained class of
family-universal models that are anomaly free without addition of new
fermions to the SM except the $e,\mu,$ and $\tau$ right-handed
neutrinos. The Abelian generator must then act on SM matter
like a linear combination of hypercharge and baryon minus lepton
number,\cite{ceg,adh} which we parameterize by an angle $\theta_X$ as
\begin{eqnarray}
Q_X= {\rm cos}\theta_X {Y^{\prime} \over 2} + 
     {\rm sin}\theta_X {(B-L)^{\prime} \over 2}.
\end{eqnarray}
If kinetic mixing occurs\cite{bh,spain} between the \zsp and \zpsp
bosons we can choose a basis for the gauge fields in which it vanishes
at the electroweak scale, resulting in a generator of the form of
equation (1) with a shifted value of $\theta_X$ --- see for
instance\cite{gm}.  Although $Y^{\prime}$ acts like the SM hypercharge
$Y$ on SM matter fields, it is a distinct $U(1)$ generator and likewise
$(B-L)^{\prime}$ is distinct from the \bmlsp generator of
$SO(10)$. In particular, we assume the \zpsp gets a large mass from an
SM singlet Higgs field with vanishing SM hypercharge but nonvanishing
\qxsp charge. This class of models was explored in \cite{adh}.  Direct
experimental bounds were extracted from LEP II data in \cite{cddt} and
CDF\cite{cdf} has obtained even stronger bounds for couplings weaker than
electroweak, $g_{Z^{\prime}}\, \ltap\, g_Z/4$. The precision
electroweak constraints and Higgs boson mass predictions for these
models were studied in \cite {flv} and \cite{mczpqx} for the EWWG data
set, without the NuTeV, M\"oller, and APV measurements.

Gauge invariance of the SM Yukawa interactions requires the SM Higgs
boson to have vanishing $(B-L)^{\prime}$ charge, so that \zzpsp mass
mixing only occurs through the $Y^{\prime}$ component of \qx. After
diagonalizing the mass matrix assuming $m_{Z^{\prime}}^2 \gg m_Z^2$ we
find\cite{mczpqx} the mass eigenstates
\begin{eqnarray}
Z&=& \cos\theta_M \, Z_0 +\sin\theta_M \, Z_0^{\prime} \\
Z^{\prime}&=& \cos\theta_M \, Z_0^{\prime} -\sin\theta_M \, Z_0.
\end{eqnarray}
The mixing angle is 
\begin{eqnarray}
\theta_M= {r \cos\theta_X \over \hat{m}_{Z^{\prime}}^2}.
\end{eqnarray}
where $r$ and $\hat{m}_{Z^{\prime}}$ are the \zpsp coupling 
constant and mass scaled to the \zsp coupling and mass, that is 
\begin{eqnarray}
r\equiv \frac{g_{Z^{\prime}}}{g_Z}
\end{eqnarray}
and 
\begin{eqnarray}
\hat{m}_{Z^{\prime}}\equiv \frac{m_{Z^{\prime}}}{m_Z}.
\end{eqnarray}
The shift in the \zsp mass generates a contribution to the oblique
parameter $T$,\cite{bh2}
\begin{eqnarray}
\alpha T_X= -{\delta m_Z^2 \over m_Z^2}
            = {r^2 \cos^2\theta_X \over \hat{m}_{Z^{\prime}}^2}. 
\end{eqnarray}

The effective $Z\overline ff$ Lagrangian can then be written as
\begin{eqnarray}
{\cal L}_f= g_Z \left(1 + {\alpha T_X \over 2}\right) g_f^{\prime}
            \overline f {\not\! Z} f
\end{eqnarray}
where $f$ represents a quark or lepton of chirality $L$ or $R$.
Since $\hat m_{Z^{\prime}}^2 \gg 1$ the mixing angle $\theta_M$ is 
very small, and to leading order in $\theta_M$ 
the $Z\overline ff$ coupling $g_f^{\prime}$ is 
\begin{eqnarray}
g_f^{\prime}=  g_f + r\theta_M q_X^f
\end{eqnarray}
where $q_X^f$ is the \qxsp charge of fermion $f$, 
\begin{eqnarray}
q_X^f = \cos\theta_X\ {y^f \over 2} +\sin\theta_X\ { b^f -  l^f\over 2}
\end{eqnarray}
and $y^f,b^f, l^f$ are respectively the weak
hypercharge, baryon number, and lepton number of fermion $f$.
The first term in equation (9) has the form of the SM $Z\overline ff$ 
coupling, 
\begin{eqnarray}
g_f= t_{3L}^f -q^f({\rm sin}^2 \theta_W^{\ell\ {\rm eff}} 
     +\delta^{\rm OB}{\rm sin}^2 \theta_W)
\end{eqnarray}
but with the oblique correction to \sinthl,
\begin{eqnarray}
\delta^{\rm OB}{\rm sin}^2 \theta_W = 
      -\frac{{\rm sin}^2 \theta_W\, {\rm cos}^2 \theta_W}
            {{\rm cos}^2 \theta_W - {\rm sin}^2 \theta_W}
            \alpha T_X.
\end{eqnarray}

For fixed $\theta_X$ the the effect of \zzpsp mixing on the
electroweak fit is determined by a single parameter which we choose to
be $T_X$. The shift in the $Z\overline ff$ coupling from the \zpsp
admixture, the second term in equation (9), is determined by
\begin{eqnarray}
\epsilon\equiv r \, \theta_M= {\alpha T_X \over  \cos \theta_X}.
\end{eqnarray}
The \chisqsp fits are obtained by varying $T_X$ in addition to the
four SM parameters, $m_t,\, \Delta\alpha^{(5)}_{\rm had}(m_Z)$,
$\alpha_S(m_Z)$, and $m_H$.  The ratio of coupling strength to mass
(the effective ``Fermi constant'') is also determined by \tx,
\begin{eqnarray}
\frac{g_{Z^{\prime}}^2}{m_{Z^{\prime}}^2}= 
              {\alpha T_X\over \cos^2 \theta_X}\frac{g_Z^2}{m_Z^2}
\end{eqnarray}
or in terms of the scaled coupling and mass,
\begin{eqnarray}
\frac{r^2}{\hat m_{Z^{\prime}}^2}=
      {\alpha T_X\over \cos^2 \theta_X}.
\end{eqnarray}
The LEP II bounds on \zpsp production\cite{cddt} can then also be
expressed in terms of \tx. Translated from the notation
of Carena {\it et al.}\cite{cddt}, who extracted the \zpsp bounds from
LEP II bounds on contact interactions, the bounds for the first
quadrant in $\theta_X$ are\cite{mczpqx}
\begin{eqnarray}
\alpha T_X \leq {1 \over (30.1 + 15.5 \tan \theta_X)^2},
\end{eqnarray}
or in terms of the scaled ratio of mass to coupling, 
\begin{eqnarray}
\frac{\hat m_{Z^{\prime}}}{r} > 30.1\cos \theta_X +15.5\sin \theta_X.
\end{eqnarray}
Stronger bounds have been obtained by CDF\cite{cdf} for \zpsp
couplings that are weaker than electroweak as discussed in some
specific cases below.

For the $Z$-pole measurements the effect of \zpsp exchange is
completely negligible but for the three low energy measurements it is
of the same order as the oblique corrections and as the \zzpsp mixing
correction to \zsp exchange.  In addition, the effect on the \ntsp
measurements includes corrections to the $Z\overline\nu\nu$
interaction which is only implicit in the notation. For instance, the
quoted measurement for the hadronic coupling $g_L^2 = g_{uL}^2
+g_{dL}^2$ implicitly includes a factor $4g_{\nu L}^2$ which is equal
to one at leading order in the SM. The result for the
nonoblique corrections, including \zpsp exchange, is 
\begin{eqnarray}
\frac{\delta g_L^2}{g_L^2}= \frac{1}{9}(1 +\tan \theta_X)
(\sin^2 \theta_W \tan\theta_X -9 +18\sin^2 \theta_W -10\sin^4 \theta_W)
\end{eqnarray}
\begin{eqnarray}
  \frac{\delta g_R^2}{g_R^2}= \frac{\sin^2 \theta_W}{9}
     [(1 +\tan\theta_X)(10 +\tan\theta_X -10\sin^2 \theta_W)
         -10 -\tan\theta_X],
\end{eqnarray}
to which we add the oblique corrections, 
\begin{eqnarray}
\delta^{\rm OB}g_L^2&=& 2\alpha T_Xg_L^2 - (1 -\frac{10}{9} {\rm
  sin}^2 \theta_W)\cdot \delta^{\rm OB}{\rm sin}^2 \theta_W\\ 
\delta^{\rm OB}g_R^2&=& 2\alpha T_Xg_R^2 + \frac{10}{9} {\rm sin}^2
\theta_W\cdot \delta^{\rm OB}{\rm sin}^2 \theta_W
\end{eqnarray}

There are extensive cancellations among the various \zpsp corrections
to both M\"oller scattering and atomic parity violation, with a simple
final result.  The sum of the \zpsp exchange correction and the \zzpsp
mixing correction to \zsp exchange cancel exactly with the oblique
wave function renormalization correction proportional to $\alpha
T_X/2$ from the prefactor in the effective Lagrangian equation
(8). The only surviving contribution is then the oblique correction to
\sinthl, equation (12).

{\it \noindent \underline{\zpsp Model Fits}}

The \zpsp fit of data set B with the best \chisqsp confidence level
consistent with the LEP II limits on $\hat m_{Z^{\prime}}/r$ is
obtained for $\theta_X= \pi/3$. It is compared with the SM fit in
table 2 and figure 1. The \chisqsp decreases by 2.9 units and the
confidence level increases from 12\% to 19\%. The improvement can be
attributed entirely to the improved fit of the \ntsp $g_L^2$
measurement, for which the \chisqsp contribution decreases by 2.7
units, from 9.3 to 6.6. More significant is the improved consistency
with the 114 GeV LEP II lower limit on the Higgs boson mass. The
central value is doubled, from 64 to 126 GeV, completely resolving the
marginal 7\% consistency level of the SM fit with the LEP II
limit. The 95\% upper limit is also doubled, from 124 to 223 GeV. The
central value for the ratio of \zpsp mass to coupling constant, $\hat
m_{Z^{\prime}}/r = 29$, is just consistent with the LEP II lower
limit,\cite{cddt} which from the inequality (16) is $\hat
m_{Z^{\prime}}/r > 28$. For very weak coupling the CDF data\cite{cdf}
for $\theta_X= \pi/3$ implies a stronger bound, $\hat m_{Z^{\prime}}/r
> 41$, which excludes the fit for $g_{Z^{\prime}} < g_Z/5$ --- see
table 5 of \cite{mczpqx}.

Figure 2 and table 3 show that fits consistent with the LEP II direct
bounds on $\hat m_{Z^{\prime}}/r$ are obtained in the entire interval
$0\leq \theta_X \leq \pi/3$ while $\pi/3 < \theta_X \leq \pi/2$ is
excluded. Table 4 shows that the fits to the EW data are also
acceptable for $0\leq \theta_X \leq \pi/3$, with central values
for \mhsp above 114 GeV and with \chisqsp confidence levels at least
as good as the SM.

\begin{figure}
\centerline{\includegraphics[width=4in,angle=90]{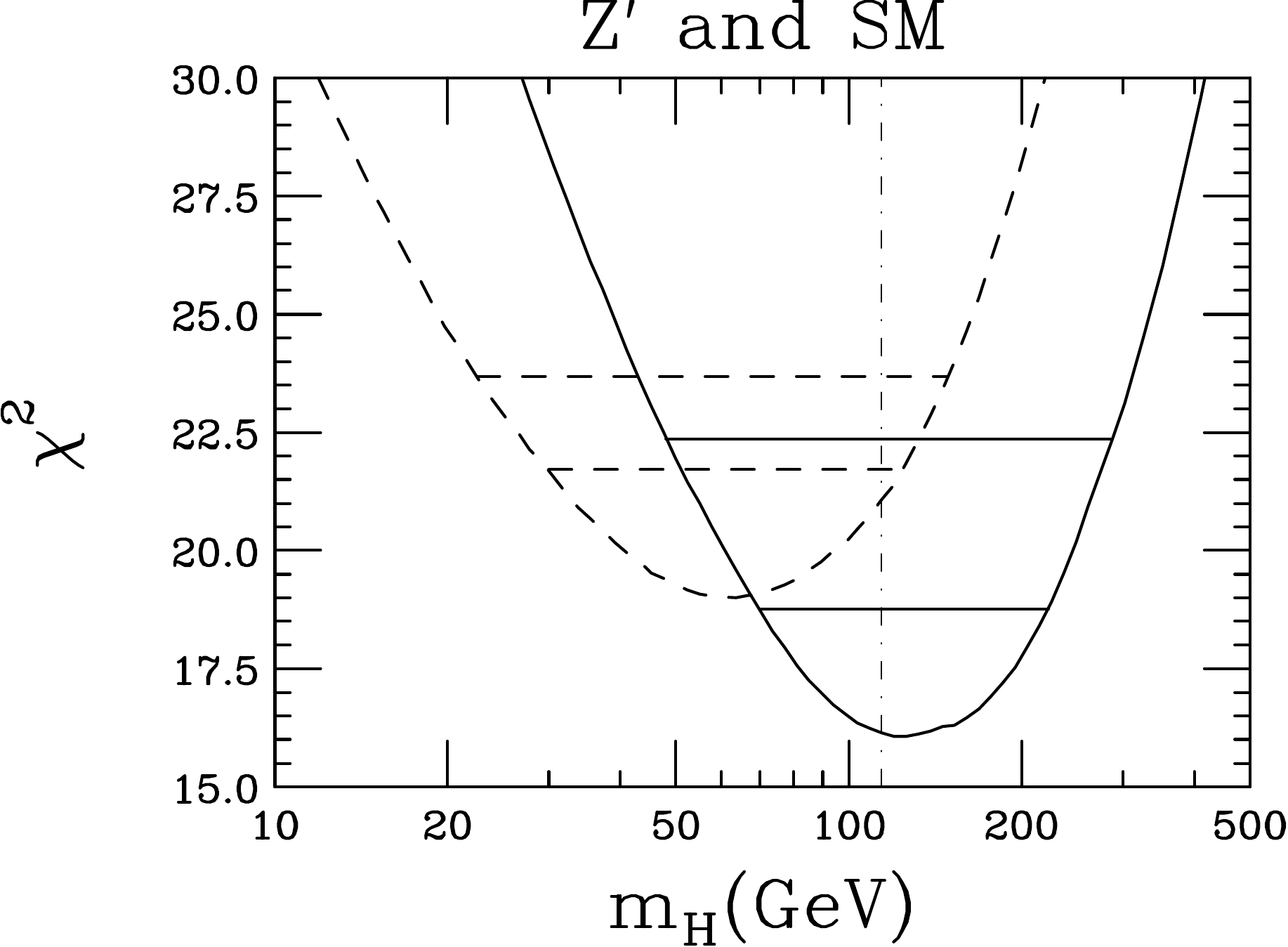}}
\caption{$\chi^2$ distributions as a function of $m_H$ for the SM fit
  (dashed lines) and the \zpsp (solid lines) fit with $\theta_X=\pi/3$
  as shown in table 2.  For each fit the lower horizontal line is the
  symmetric 90\% confidence interval and the upper horizontal line is
  the Bayesian 95\% confidence interval defined in the text. The
  vertical dash-dot line indicates the LEP II direct lower limit on
  \mh.  }
\label{fig1}
\end{figure}

\begin{table}
\begin{center}
\vskip 12pt
\begin{tabular}{c|c|cc|cc}
\hline
\hline
 &Experiment& {\bf SM} & Pull &{\bf \zp} & Pull \\ 
\hline
$A_{LR}$ & 0.1513 (21) & 0.1494  & 0.9 &0.1511 &0.08  \\
$A_{FB}^l$ & 0.01714 (95) &0.1674  &0.4 &0.0712 &0.02    \\
$A_{e,\tau}$ & 0.1465 (32) & 0.1494 & -0.9 & 0.1511 &-1.5 \\
$m_W$ & 80.398 (25) & 80.391 & 0.3 & 80.370 & 1.1 \\
$\Gamma_Z$ & 2495.2 (23) & 2496.1 & -0.4  & 2495.5 &-0.1 \\
$R_l$ & 20.767 (25) &20.743  & 1.0 & 20.753 & 0.6 \\
$\sigma_h$ & 41.540 (37) & 41.479 &1.7 & 41.491 & 1.3   \\
$R_b$ & 0.21629 (66) & 0.21584 &0.7 & 0.21582 & 0.7 \\
$R_c$ & 0.1721 (30) & 0.1722 &-0.04 & 0.1723 & -0.08 \\
$A_b$ & 0.923 (20) & 0.935 &-0.6 & 0.935 & -0.6 \\
$A_c$ & 0.670 (27) &  0.669 & 0.04 & 0.669 &0.03 \\
$g_L^2$& 0.30005 (137)& 0.30423& -3.1 & 0.30357 & 2.6 \\
$g_R^2$ &0.03076 (11)& 0.03004& 0.7 & 0.03022 & 0.5  \\
$x_W(ee)$& 0.23339 (140)& 0.23122& 1.55 & 0.23147 &1.4   \\
$x_W(Cs)$ &0.22939 (190)& 0.23122& -1.0 & 0.23147 & -1.1   \\
$m_t$ & 172.6 (1.4) &172.3  &0.2 & 172.3 &0.2    \\
$\Delta \alpha_5(m_Z)$ & 0.02758 (35) &0.02754 & 0.1 & 0.2761 & -0.09   \\
$\alpha_S(m_Z)$ &0.118  & & 0.120 &    \\
\hline
$m_H$ & & 64 & & 126 &  \\
CL$(m_H > 114)$ && 0.07 & & 0.60 &   \\
$m_H(95\%)$&&124& & 223 &  \\
\hline
$T_X$ && && 0.037& \\
$\hat m_{Z^{\prime}}/r$ && && 29 &\\
\hline
$\chi^2$/dof& & 19.0/13 & & 16.1/12 &   \\
CL($\chi^2)$ &&0.12 & & 0.19 &   \\
\hline
\hline
\end{tabular}
\end{center}
\caption{The SM fit of data set (B) compared to the \zpsp fit with 
$\theta_X=\pi/3$.}
\end{table}

\begin{figure}
\centerline{\includegraphics[width=7in,angle=0]{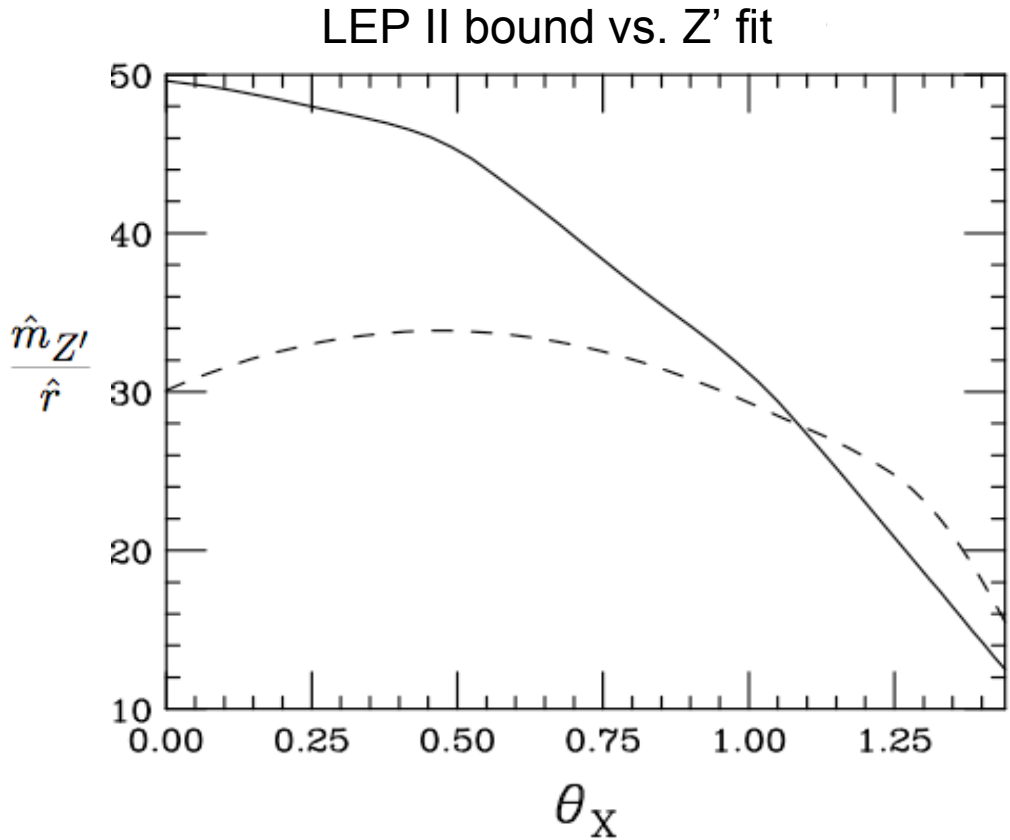}}
\caption{$\hat m_{Z^{\prime}}/r$ as a function of $\theta_X$ for the 
\zpsp fits (solid line) compared with the LEP II 95\% lower limits given by 
equation (17).
}
\label{fig2}
\end{figure}

\begin{table}
\begin{center}
\vskip 12pt
\begin{tabular}{c||cc|c}
\hline
\hline
Model & $\hat m_{Z^{\prime}}/r$ & $\hat m_{Z^{\prime}}/r$[LEP II] 
        & $m_{Z^{\prime}}|_{r=1}$(TeV) \\  
\hline
$\theta_X=0 $&50&$>\, 30$ & 4.6 \\
$\theta_X=\pi/12 $& 48 &$>\, 33$ &4.4 \\
$\theta_X=\pi/6$&45 &$>\, 34$ & 4.1 \\
$\theta_X=\pi/4 $ &37 &$>\, 32$ & 3.4 \\
$\theta_X=\pi/3 $ &29 &$>\, 28$ & 2.6\\
\hline
\hline
\end{tabular}
\end{center}
\caption{The value of $\hat m_{Z^{\prime}}/r$ from the electroweak fit
  compared with the LEP II lower bound. The last column shows the
  \zpsp mass if $g_{Z^{\prime}}=g_Z$, i.e., if $r=1$.}
\end{table}

The entire \qxsp parameter space is spanned by the first two quadrants
of $\theta_X$, which are equivalent to the third and fourth quadrants
up to the sign of $g_{Z^{\prime}}$.  In the second quadrant, $\pi/2
\leq \theta_X \leq\pi$, there are no \zpsp fits which improve on
either the SM \mhsp prediction or the \chisqsp confidence level. In
particular, for the ``$\chi$'' boson of SO(10) which couples to
$T_{3R}-(B-L)/2$, corresponding to $\theta_X = {\rm arctan(-2)}$, zero
mixing is preferred and the \chisqsp is the same as in the SM fit.

Figure 1 and table 4 display ``Bayesian'' 95\% confidence limits for
\mhsp which differ from the conventionally quoted 95\% CL frequentist
upper limits that are also shown. Anticipating the discovery of a
Higgs boson at the LHC, today's Bayesian fit will become the
conventional frequentist fit when the Higgs boson is discovered. Since
we will then not scan on \mhsp there is one additional degree of
freedom. The Bayesian 95\% interval is the domain of \mhsp with
confidence level $\geq 5\%$ for fits with \mhsp fixed and indicates
the allowed range in \mhsp {\em after} the Higgs boson is
discovered. In contrast, the usual frequentist 95\% upper limit (the
maximum of the symmetric 90\% confidence interval) uses the best fit
with \mhsp scanned as a free parameter to estmate the likelihood that
the Higgs boson {\em will be discovered} within the given range of
values. Table 4 shows that the Bayesian limits lie above the
frequentist limits and that they are also increased in
the \zpsp models relative to the SM.

The \zpsp fits to data set A, which includes the hadronic asymmetries
\afbbsp and \afbc, are not shown, because they are little changed with
respect to the SM. For most of the $\theta_X$ domain they are not
changed at all. In the most favorable fit \chisqsp only decreases 
from 28.4 to 28.0, with a confidence level of 0.014, less than the
0.019 confidence level of the SM fit, and with only a modest increase
in the Higgs boson mass, from 94 to 130 GeV.

As observed in \cite{gambinoetal} the results presented by
\ntsp\cite{nutev} for \sinthwsp and $g_{L,R}^2$ are sensitive to the
assumption that the nucleon strange quark sea is symmetric between $s$
and $\overline s$ quarks.  The \ntsp collaboration has since found a
$1.4\sigma$ indication of an asymmetry in the momentum carried by
strange quarks, $S^-=+0.0020 (14)$.\cite{ntssbar} The collaboration
has not revised their value for \sinthwsp based on this result, but
they do state that an asymmetry of +0.007 would make their measurment
of \sinthwsp consistent with SM prediction. Using this statement and
taking the new $S^-$ measurement at face value, the PDG\cite{pdg} has
estimated revised values for \sinthw, $g_L^2$, and $g_R^2$, which we
use in the alternative fits shown in figure 3 and tables 5 and
6.\footnote{The anomaly in \sinthwsp is 0.2277 - 0.2227= 0.0050 so the
  shift from $S^-=+0.0020 (14)$ is 2/7 of the anomaly or $-0.0014
  (10)$. Combining errors in quadrature then gives the values quoted
  by the PDG.}  With these estimates the disagreement with the SM is
reduced to a $2\sigma$ effect.

As can be seen by comparing the SM fits of tables 1 and 5, the effect
of the revised estimate of the \ntsp results on the global fit to data
set B is to raise the confidence level from 0.12 to 0.35, while the
inconsistency with the LEP II constraint on \mhsp remains: the central
value is $m_H=58$ GeV and the confidence level for \mhsp above 114 GeV
is 6\%. Again the \zpsp models can resolve the inconsistency with the
LEP II lower bound on \mhsp while maintaining the \chisqsp confidence
level of the fit. In particular, for $\theta_X = \pi/3$, shown in
figure 3 and table 5, the central value doubles, to 109 GeV, and the
CL for $m_H > 114$ GeV increases to 45\%.  All \zpsp models in the
domain $0\leq\theta_X\leq\pi/3$ increase the \mhsp
prediction into the LEP II allowed region, with acceptable fits to the
precision data and with values of $m_{Z^{\prime}}/g_{Z^{\prime}}$
consistent with the LEP II lower bounds as shown in tables 6 and 7.

\begin{table}
\begin{center}
\vskip 12pt
\begin{tabular}{c||c|cc|cc|cc}
\hline
\hline
Model & $T_X$&\chisq &CL(\chisq) &\mh &CL($m_H>114)$ & 
   $m_H^{95\%}({\rm Freq.})$ & $m_H^{95\%}({\rm Bayes})$ \\
\hline
SM & &19.0 & 0.12 & 64 & 0.07 & 124 & 149\\
\hline
$\theta_X=0 $&0.052&17.9 &0.12 &120 &0.56 & 215 &249 \\
$\theta_X=\pi/12 $&0.052&17.4 &0.14 &126 &0.58 &224 &266 \\
$\theta_X=\pi/6$&0.048&16.9 &0.15 &126 &0.59 &223 &281 \\
$\theta_X=\pi/4 $&0.046&16.5 &0.17 &126 &0.60 &230 &300 \\
$\theta_X=\pi/3 $&0.037&16.1 &0.19 &126 &0.60 &223 &288 \\
\hline
\hline
\end{tabular}
\end{center}
\caption{Best fit value of \txsp, \chisqsp confidence level, and \mhsp
  predictions for SM and \zpsp fits to data set B.}
\end{table}

\begin{figure}
\centerline{\includegraphics[width=4in,angle=90]{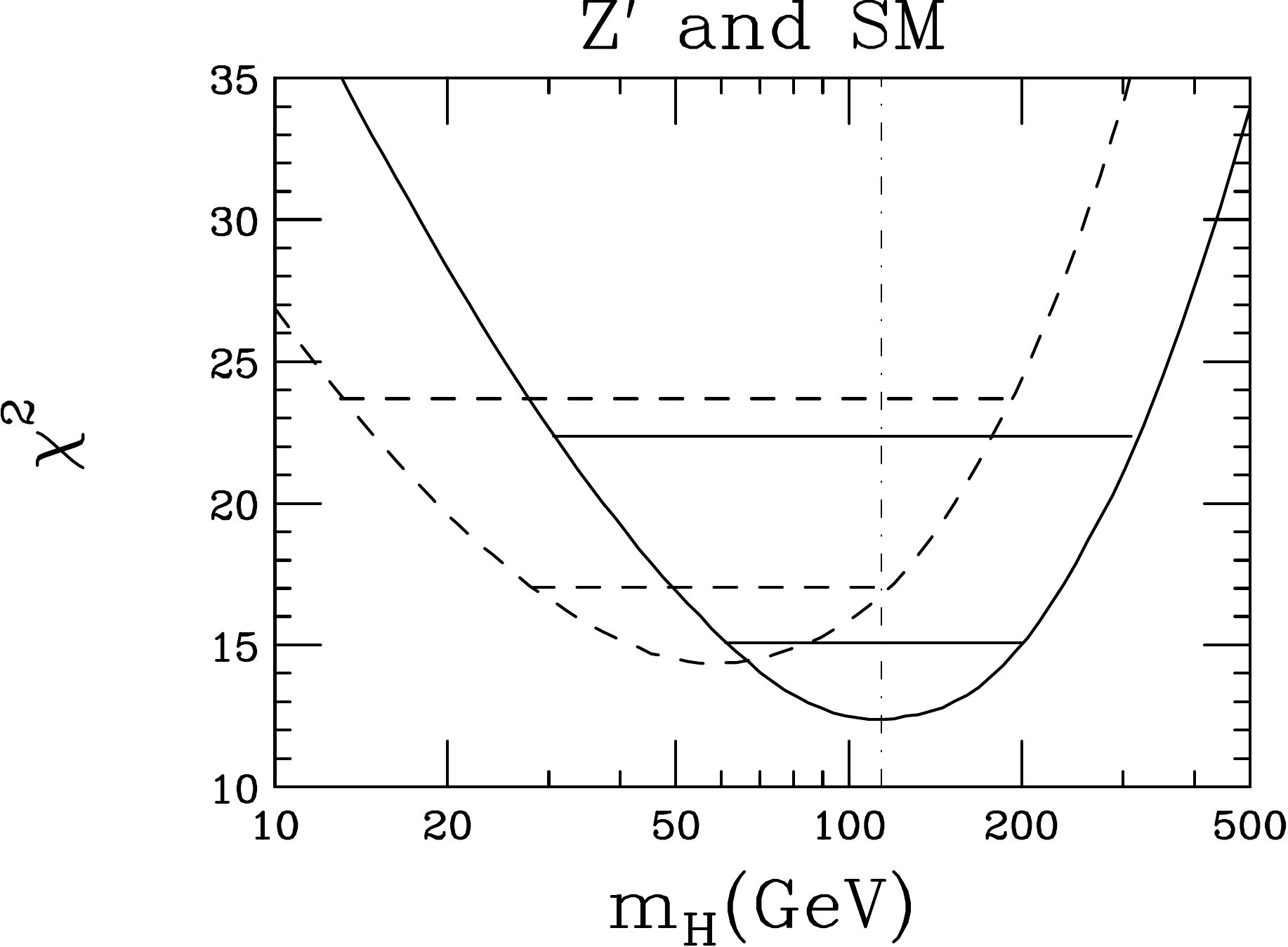}}
\caption{$\chi^2$ distributions as a function of $m_H$ for the SM
  (dashed lines) and \zpsp (solid lines) fits shown in table 5, with PDG
  values for $g_L^2$ and $g_R^2$.  For each fit the lower horizontal
  line is the symmetric 90\% confidence interval and the upper
  horizontal line is the Bayesian 95\% confidence interval defined in
  the text. The vertical dash-dot line indicates the LEP II direct
  lower limit on \mh.  }
\label{fig3}
\end{figure}

\begin{table}
\begin{center}
\vskip 12pt
\begin{tabular}{c|c|cc|cc}
\hline
\hline
 &Experiment& {\bf SM} & Pull &{\bf \zp} & Pull \\ 
\hline
$A_{LR}$ & 0.1513 (21) & 0.1498  & 0.7 &0.1510 &0.1  \\
$A_{FB}^l$ & 0.01714 (95) &0.1682  &0.3 &0.0710 &0.05    \\
$A_{e,\tau}$ & 0.1465 (32) & 0.1498 & -1.0 & 0.1510 &-1.4 \\
$m_W$ & 80.398 (25) & 80.396 & 0.08 & 80.376 & 0.9 \\
$\Gamma_Z$ & 2495.2 (23) & 2496.2 & -0.4  & 2495.5 &-0.1 \\
$R_l$ & 20.767 (25) &20.744  & 0.9 & 20.750 & 0.7 \\
$\sigma_h$ & 41.540 (37) & 41.479 &1.7 & 41.491 & 1.3   \\
$R_b$ & 0.21629 (66) & 0.21584 &0.7 & 0.21581 & 0.7 \\
$R_c$ & 0.1721 (30) & 0.1722 &-0.05 & 0.1723 & -0.07 \\
$A_b$ & 0.923 (20) & 0.935 &-0.6 & 0.935 & -0.6 \\
$A_c$ & 0.670 (27) &  0.669 & 0.04 & 0.669 &0.03 \\
$g_L^2$& 0.3010 (15)& 0.3043& -2.2 & 0.3037 & 1.8 \\
$g_R^2$ &0.03080 (11)& 0.03003& 0.7 & 0.03020 & 0.5  \\
$x_W(ee)$& 0.23339 (140)& 0.23117& 1.6 & 0.23144 &1.4   \\
$x_W(Cs)$ &0.22939 (190)& 0.23117& -0.9 & 0.23144 & -1.1   \\
$m_t$ & 172.6 (1.4) &172.3  &0.2 & 172.3 &0.2    \\
$\Delta \alpha_5(m_Z)$ & 0.02758 (35) &0.02754 & 0.1 & 0.2768 & -0.3   \\
$\alpha_S(m_Z)$ &0.118  & & 0.120 &    \\
\hline
$m_H$ & & 58 & & 109 &  \\
CL$(m_H > 114)$ && 0.06 & & 0.45 &   \\
$m_H(95\%)$&&118& & 201 &  \\
\hline
$T_X$ & & && 0.033& \\
$\hat m_{Z^{\prime}}/r$ &&&& 31 &\\
\hline
$\chi^2$/dof& & 14.4/13 & & 12.4/12 &   \\
CL($\chi^2)$ &&0.35 & & 0.42 &   \\
\hline
\hline
\end{tabular}
\end{center}
\caption{SM and \zpsp ($\theta_X=\pi/3$) fits to data set B with 
PDG values for $g_L^2$ and $g_R^2$.}
\end{table}

\begin{table}
\begin{center}
\vskip 12pt
\begin{tabular}{c||c|cc|cc|cc}
\hline
\hline
Model & $T_X$& \chisq &CL(\chisq) &\mh &CL($m_H>114)$ & 
   $m_H^{95\%}({\rm Freq.})$ & $m_H^{95\%}({\rm Bayes})$ \\
\hline
SM & &14.3 & 0.35 & 58 & 0.06 & 118 & 192\\
\hline
$\theta_X=0 $&0.043&13.7 &0.32 &104 &0.42 &189  &285 \\
$\theta_X=\pi/12 $&0.044&13.3 &0.35 &109 &0.42 &197 &300 \\
$\theta_X=\pi/6$&0.043&13.0 &0.37 &114 &0.50 &203 &312 \\
$\theta_X=\pi/4 $&0.039&12.7 &0.39 &114 &0.50 &202 &317 \\
$\theta_X=\pi/3 $&0.033&12.4 &0.42 &109 &0.45 &201 &320 \\
\hline
\hline
\end{tabular}
\end{center}
\caption{Best fit value of \txsp, \chisqsp confidence level, and \mhsp
  predictions from SM and \zpsp fits to data set B with the PDG
  estimates for $g_L^2$ and $g_R^2$.}
\end{table}

\begin{table}
\begin{center}
\vskip 12pt
\begin{tabular}{c||cc|c}
\hline
\hline
Model & $\hat m_{Z^{\prime}}/r$ & $\hat m_{Z^{\prime}}/r$[LEP II] 
        & $m_{Z^{\prime}}|_{r=1}$(TeV) \\  
\hline
$\theta_X=0 $&55&$>\, 30$ & 5.0 \\
$\theta_X=\pi/12 $& 52 &$>\, 33$ &4.7 \\
$\theta_X=\pi/6$&47 &$>\, 34$ & 4.3 \\
$\theta_X=\pi/4 $ &40 &$>\, 32$ & 3.6 \\
$\theta_X=\pi/3 $ &31 &$>\, 28$ & 2.8\\
\hline
\hline
\end{tabular}
\end{center}
\caption{The value of $\hat m_{Z^{\prime}}/r$ from the fit with PDG
  estimates for $g_{L,R}^2$ compared with the LEP II lower bound.
  The last column shows the \zpsp mass if
  $g_{Z^{\prime}}=g_Z$, i.e., if $r=1$.}
\end{table}

For data set A with the PDG estimates of $g_L^2$ and $g_R^2$, the SM
fit yields $\chi^2/N= 24.2/15$ with a still marginal confidence level
of 6\%. The Higgs mass prediction is acceptable, at $m_H= 85$ GeV,
with 30\% likelihood for $m_H > 114$ GeV. In most cases the \zpsp fits
prefer zero mixing, and in all cases the effect on \chisqsp and \mhsp 
is negligible.
\newpage
{\it \noindent \underline{Conclusion}}

We have extended a previous study of a simple class of
\zpsp models to include three low energy measurements in addition to
the EWWG set of $Z$-pole observables. In the earlier study we found
that the \zpsp models had little effect on the \chisqsp or on the
central value of \mhsp but that they did significantly extend the
upper limit on \mhsp for the data set without \afbbsp and \afbc,
increasing $CL(m_H > 114 {\rm GeV})$ from 2\% in the SM to as much as
30\% in the \zpsp models.  With the inclusion of the low energy
measurements and, in particular, the \ntsp data, the effect of the
\zpsp models is more pronounced. Both the central value and the upper
limit of the Higgs boson mass are increased significantly, by as much
as a factor two, and the \chisqsp confidence levels are modestly
improved. If underestimated systematic error proves to be the
explanation of the $3.2\sigma$ difference in the SM determination of
\sinthlsp from the leptonic and hadronic asymmetry measurements, then
the \zpsp models discussed here can resolve the resulting problematic
prediction for the Higgs boson mass, providing completely satisfactory
fits to the precision EW data. Though it remains to be confirmed by a
careful study, we expect that the \zpsp models favored by the fits can
be excluded or confirmed at the LHC.

\bigskip

\noindent{\small This work was supported in part by the Director, Office of
Science, Office of High Energy and Nuclear Physics, Division of High
Energy Physics, of the U.S. Department of Energy under Contract
DE-AC02-05CH11231}

\end{document}